\documentclass[12pt]{article}
\usepackage{graphicx}
\usepackage{amssymb}
\usepackage{scicite}

\usepackage{amsmath}
\usepackage{url}

\newcommand\TF{\mathit{TF}}
\newenvironment{sciabstract}{%
\begin{quote} \bf}
{\end{quote}}

\title{Is coaching experience associated with effective use of timeouts in basketball?}
\author{Serguei Saavedra$^{1,2}$\footnote{To whom correspondence should be addressed. E-mail: s-saavedra@northwestern.edu}, Satyam Mukherjee$^{1,3}$, and James P. Bagrow$^{1,4}$\\ 
\\      
\\$^1$Northwestern Institute on Complex Systems\\ Northwestern University\\ Evanston, Illinois, 60208, USA\\
\\$^2$Northwestern University Clinical and Translational Sciences Institute\\ Northwestern University\\ Chicago, IL 60611\\
\\$^3$Kellogg School of Management\\ Northwestern University\\ Evanston, Illinois, 60208, USA\\
\\$^4$Engineering Sciences and Applied Mathematics\\ Northwestern University\\ Evanston, Illinois, 60208, USA\\  
 }

\date{}

\begin{document}
\maketitle
\baselineskip=8.5mm
 
\vspace{0.4 in}

\newpage
 
\begin{sciabstract}
Experience is an important asset in almost any professional activity. In basketball, there is believed to be a positive association between coaching experience and effective use of team timeouts. Here, we analyze both the extent to which a team's change in scoring margin per possession after timeouts deviate from the team's average scoring margin per possession---what we called timeout factor, and the extent to which this performance measure is associated with coaching experience across all teams in the National Basketball Association over the 2009-2012 seasons. We find that timeout factor plays a minor role in the scoring dynamics of basketball. Surprisingly, we find that timeout factor is negatively associated with coaching experience. Our findings support empirical studies showing that, under certain conditions, mentors early in their careers can have a stronger positive impact on their teams than later in their careers.
\end{sciabstract}

\vspace{0.2 in}


Experience is one of the most important parameters used to evaluate someone's potential performance and mentorship skills at almost any professional activity \cite{Kram}. However, broad empirical facts regarding the link between experience and performance are only now emerging due to the complication of gathering field data or constructing experiments \cite{Guimera,Neiman}. In fact, research has shown that in academia, mentors early in their careers can have a stronger positive impact on prot\'{e}g\'{e}s than later in their careers \cite{Malmgren}. In basketball, there are many important aspects that can characterize the experience of a coach such as game strategy, motivation skills, effective use of timeouts, among others \cite{Mace,Skinner}. For many years, there has been a common belief that team timeouts (i.e., strategic breaks) can impact, positively or negatively, scoring in basketball \cite{Mace,Roane}. Typically, coaches call timeouts to change negative momentum, to rest or change players, to inspire morale, to discuss plays, or to modify their game strategy \cite{Mace,Roane}. Indeed, previous research has shown that timeouts can change the momentum of a game over short periods of time \cite{Mace,Roane}. However, during timeouts, both teams have the opportunity to take advantage of this strategic break, and it is currently unknown whether timeouts can actually change the scoring dynamics in basketball and whether coaching experience is associated with effective use of timeouts. Importantly, a wealth of data are available for sports \cite{Duch,Saavedra,Skinner2,Radicchi}, whose unambiguous performance measures provide an excellent opportunity to investigate untested ideas such as the timeout factor. Here, we quantified both the extent to which scoring dynamics after timeouts deviate from what would be expected by chance, and the extent to which team performance after timeouts is associated with coaching experience across all teams in the National Basketball Association (NBA).

To investigate the timeout factor, we used actual time series of scores and all timeouts called in more than 3000 games over the 2009-2012 seasons of the NBA. These time series were collected directly from the NBA website \cite{NBA}, where there are detailed play-by-play records for each game. We defined the timeout factor as the extent to which a team's change in scoring margin per possession after timeouts deviate from the team's average scoring margin per possession in each season. Each opportunity to score points in the game is called a possession, which lasts from the time a team obtains the basketball until the time their opponent gains possession of the basketball. The number of possessions works as a reliable measure in basketball to standardize the points scored during any interval of time \cite{Kubatko}. 

Mathematically, the timeout factor is defined as $\TF_i(n)=\rho_i(n)-\langle \rho_i^*\rangle$, where $\rho_i(n)=\left(nT_i\right)^{-1}\sum_{j=1}^{T_i}\Delta_{i,j}(n)$ is the team's change in scoring margin per possession after timeouts; and $\langle \rho_i^*\rangle=\left(m_{i,k}G_i\right)^{-1}\sum_{k=1}^{G_i} \delta_{i,k}$ is the team's average scoring margin per possession in each season. Here, $n$ is the number of possessions considered after experiencing a timeout (i.e., a timeout called either by team $i$ or the opposing team), $T_i$ is the total number of timeouts experienced by team $i$ across all quarters and games, $\Delta_{i,j}(n)$ is the change in scoring margin between the scoring margin over $n$ possessions after timeout $j$ was called and the scoring margin at the time when the timeout $j$ was called, $G_i$ is the total number of games played by team $i$, $m_{i,k}$ is the total number of possessions in game $k$, and $\delta_{i,k}$ is the final scoring margin in game $k$. Scoring margins are the difference in points between team $i$ and the opposing team at a given time. Note that the numbers of possessions $n$ and $m$ are used, respectively, to capture the scoring dynamics over a short period of time and during the entire game. The higher the timeout factor $\TF_i(n)$, the higher the team performance after timeout relative to the team's average.

\subsection*{Results}

First, to analyze whether the timeout factor plays a significant role in the scoring dynamics of basketball \cite{Redner,Skinner2,Yaari}, we used a Monte Carlo approach and compared the observed timeout factors $\TF_i(n)$ over a fixed number of possessions $n$ to the timeout factor $\TF^*_i(n)$ that would be expected by chance if timeouts were called randomly during the game. To calculate $\TF^*_i(n)$, we took the actual time series of scores for each game, randomly placed the timeouts preserving the number of timeouts of each quarter, then calculated the timeout factor as normal. The statistical significance is defined as $z_i = (\TF_i(n) - \langle \TF^*_i(n)\rangle)/\sigma_{\TF^*_i(n)}\,$, where $\langle \TF^*_i(n) \rangle$ and $\sigma_{\TF^*_i(n)}$ are the average and standard deviation of the expected timeout factors across an ensemble of 1000 random replicates within which the timeouts in each game have been randomized. Actual timeouts are more likely to occur during certain game times, for example near the end of each quarter (Fig.\ 1); in our randomizations we also preserved the observed distribution of timeouts per minute. This process controls for both the number and timing of timeouts experienced by each team. Here, $-2<z_i<2$ are considered non-significant timeout factors.

We found that timeout factor plays only a minor role in the scoring dynamics of basketball. Note that if one considers a binomial model $B(90,0.05)$ over $90$ cases (we considered 30 unique teams in each of the three seasons), timeouts would prevail as a significant variable across the NBA if at most $8$ cases showed no significant performances. However, Fig.\ 2 shows that, in all the observed number of possessions $n$ after timeout, the timeout factor falls within the non-significant range in more than 74 cases. The number of non-significant cases increases as the number of possessions increases. Importantly, after the third possession, we found that the number of non-significant cases is greater than 84, meaning that the opposite null hypothesis $B(90,0.95)$ that timeouts play no role in the scoring dynamics cannot be rejected. This suggests that the timeout factor may only last until the third possession after timeout. Similarly, the number of non-significant cases is still higher than expected by chance when we considered the fourth quarter alone. These results held when we analyzed only the timeouts called by either team separately, where we quantified the effect on a team of only the timeouts that the team itself called, and removing the effect of timeouts that the opposing team called. These findings support the idea that scoring dynamics can be explained by simple random processes \cite{Redner}, and suggest that external strategies, such as timeouts, play a minor role in basketball. Nevertheless, the question remains, are teams with higher timeout factors coached by high-experienced coaches?

To answer these questions, we quantified the association between timeout factor $\TF_i(n)$ and coaching experience. The latter was evaluated by the number of years that each coach had been head coach in the NBA prior to our observation period. Results held when we used the number of years that each coach had had any coaching experience (e.g. assistant coach or college basketball). These data were collected from individual coaches' profiles in Wikipedia \cite{Wiki}. Importantly, we found a negative ($-0.29<r<-0.24$) and significant ($p<0.015$ with Markov hypothesis testing) association between coaching experience and timeout factor for the first, second and third possession after timeout (Fig.\ 3). In 4 or more possessions the association is negative but non-significant. Figure 3 shows that teams with coaches early in their careers display on average positive timeout factors, while teams with high-experienced coaches display on average negative timeout factors. Note that coaches with more than 20 years of experience show on average comparable low timeout factors. Our results held even when we removed any potential outliers. Interestingly, if we only consider coaches with more than 1 year of experience, the negative correlation for the first, second and third possessions becomes even stronger ($-0.45<r<-0.41$), suggesting that the first year can be one of the most difficult for coaches. While a complete determination of the drivers of these patterns is beyond our analysis, one possible way to account for this difference between coaches is that coaches early in their careers might be using more risky strategies and in consequence feature a higher-than-average variance in outcomes \cite{Neiman,Skinner}.  Additionally, we found no significant association ($p>0.05$ using Markov hypothesis testing) between timeout factor and team payroll \cite{payroll}, which suggests that richer teams are not particularly better at capitalizing on timeouts.

\subsection*{Discussion}

In line with previous research that has shown that some common beliefs such as the ``hot-hand'' factor are not true in basketball \cite{Gilovich,Yaari,Wardrop}, here using a Monte Carlo approach and in the absence of other evidence and calculations, we have statistically demonstrated that timeouts play a minor role in basketball. While both teams may use timeouts to restore players' physical and mental fatigue, our results reveal that timeouts should be considered neither an advantage nor a detriment to the scoring dynamics in basketball. Nevertheless, we found that on average teams with coaches early in their careers benefit relatively more from timeouts than teams with high-experienced coaches. Interestingly, in academia, early in their careers mentors have also been found to have a significantly positive impact on their prot\'{e}g\'{e}s; while late in their careers mentors can have a significantly negative impact \cite{Malmgren}. While experience is important for many other different activities within an organization or profession, these findings suggest that not only in academia but also in sports, people early in their careers can have a strong positive effect on others.

\clearpage
\subsection*{Acknowledgments}
We would like to thank Peter Mucha and an anonymous referee for useful comments on a previous manuscript. We also thank the Kellogg School of Management, Northwestern University, the Northwestern Institute on Complex Systems (NICO), and the Army Research Laboratory under Cooperative Agreement W911NF-09-2-0053 for financial support. SS also thanks NUCATS grant UL1RR025741 and CONACYT.

\subsection*{Author Contributions}
SM extracted the data. SS, SM and JB analyzed the data and wrote the main manuscript text. SS prepared figures 1-3. All authors reviewed the manuscript.

\subsection*{Additional Information}
The authors declare no competing financial interests.

\clearpage
\medskip

\renewcommand{\baselinestretch}{1.5}
{\small
\bibliographystyle{Science}

\clearpage

\begin{figure}[tt]
\centerline{\includegraphics*[width=4in]{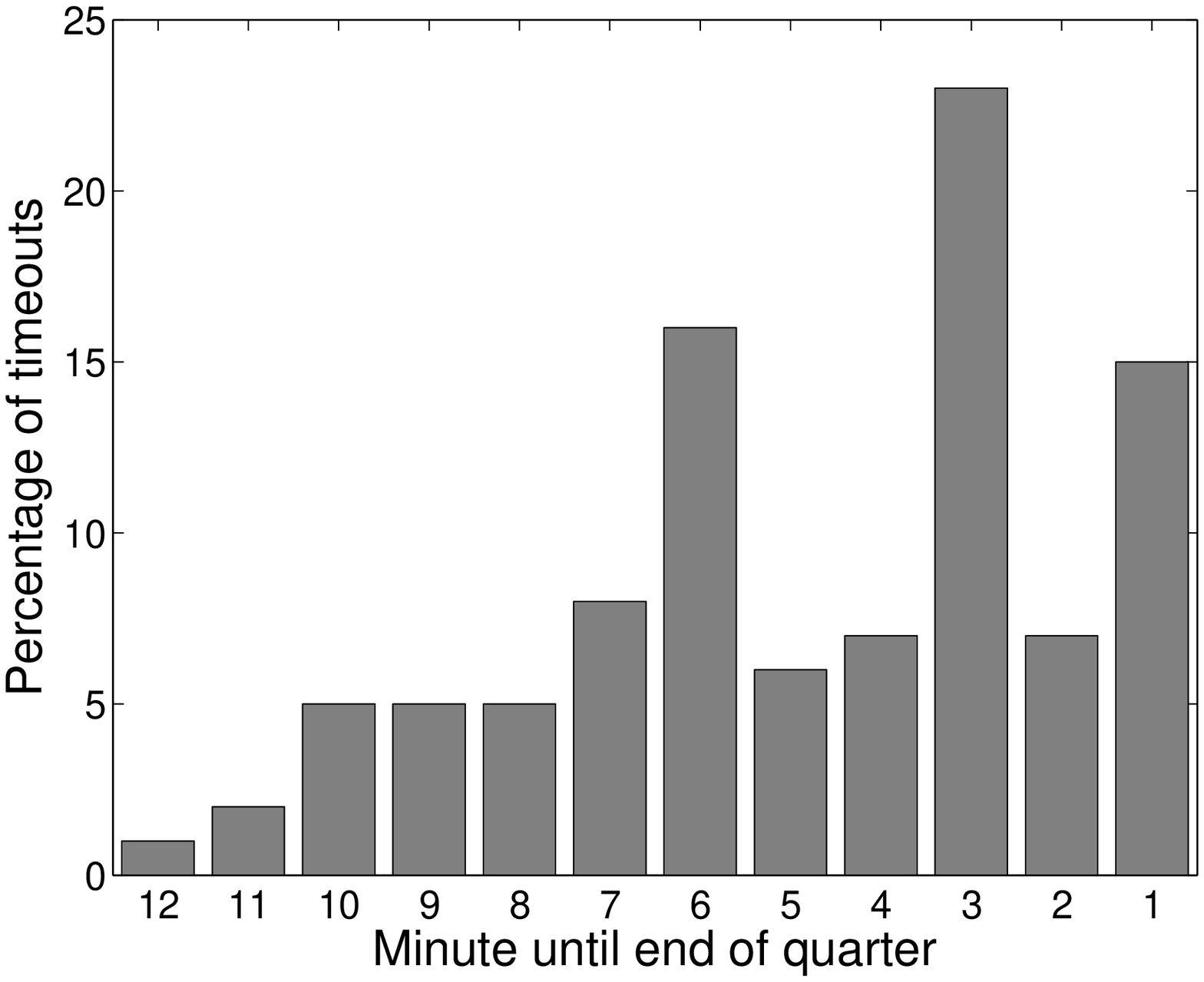}}
\caption{Distribution of timeouts. The figure shows the observed distribution of timeouts across the three seasons. Actual timeouts are more likely to occur near the end of each quarter. In our randomizations we preserved the observed distribution of timeouts per minute.}
\label{fig1}
\end{figure}

\clearpage

\begin{figure}[tt]
\centerline{\includegraphics*[width=5in]{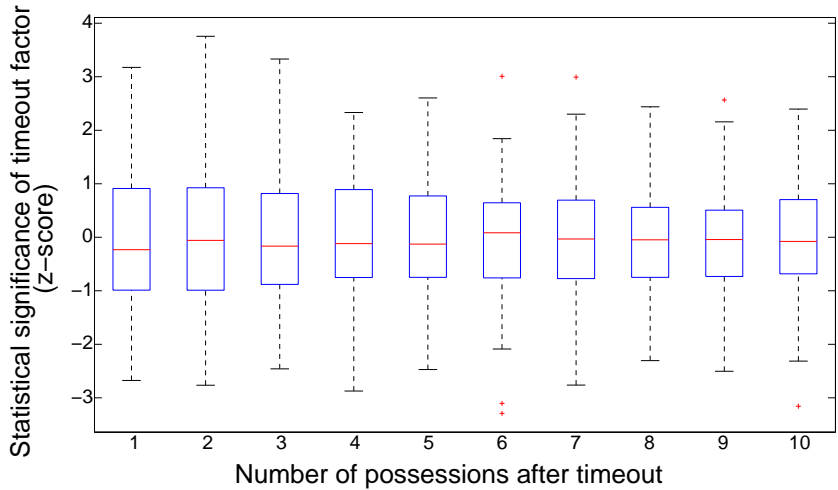}}
\caption{Statistical significance of timeout factor. The distribution (boxplots) of statistical significance ($z$-score) of timeout factor across all the NBA teams as a function of number of possessions $n$ after timeout. Note that we considered 30 unique teams in each of the three seasons, which generated 90 cases. The timeout factor falls within the non-significant range $-2<z<2$ in 74, 77 and 79 cases out of 90 in the first, second and third possessions, respectively. In 4 or more possessions, the number of non-significant cases is always higher than 84. Note that if one considers a binomial model $B(90,0.05)$ over $90$ cases, timeouts would prevail as a significant variable across the NBA if at most $8$ cases showed no significant values.}
\label{fig2}
\end{figure}

\clearpage

\begin{figure}[tt]
\centerline{\includegraphics*[width=4.5in]{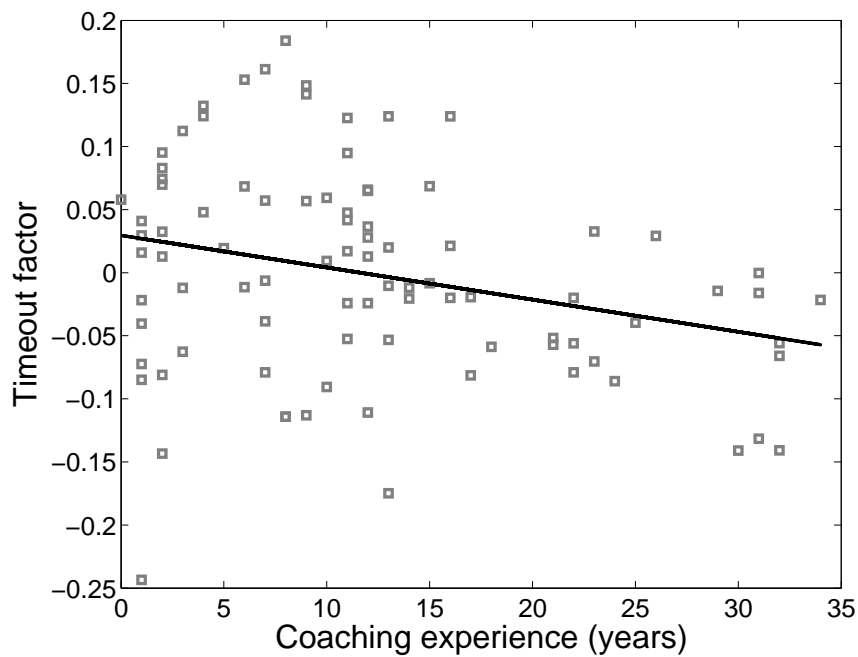}}
\caption{Association between timeout factor and coaching experience. The figure shows a negative ($r=-0.25$) and significant ($p=0.0076$, using Markov hypothesis testing) relationship between coaching experience and timeout factor $\TF_i(1st)$ over the first possession after timeout. Similar patterns were found for two and three possessions. In 4 or more possessions the association is negative but non-significant. Solid line corresponds to the best linear fit. Coaching experience was evaluated by the number of years that each coach had been head coach in the NBA prior to our observation period (since each coach debuted). Note that coaches with more than 20 years of experience show on average negative timeout factors. Results held when we used the number of years that each coach had had any coaching experience (e.g. assistant coach or college basketball). Interestingly, if we only consider coaches with more than 1 year of experience, the negative correlation becomes even stronger ($r=-0.42$), suggesting that the first year can be one of the most difficult for coaches.}
\label{fig3}
\end{figure}

\end{document}